\documentclass[sigconf]{acmart}

\usepackage{bm}
\usepackage{mathtools}
\usepackage{multirow}
\usepackage{color}
\usepackage{enumitem}
\usepackage{balance}
\usepackage[utf8]{inputenc}
\usepackage{xspace}
\usepackage{amsmath}

\newcommand{\alphaval}[2]{{\small $p\,#1\,#2$}}

\newcommand{\wilcoxon}[3]{{\small $Z=#1$, \alphaval{#2}{#3}}}

\newcommand{\cmt}[4]{\ifx\DRAFT\undefined\else\colorbox{#3}{\textcolor{#4}{\small{\textsf{[\textbf{#1}: #2]}}}}\fi}
\newcommand{\ph}[1]{\ifx\DRAFT\undefined\else\colorbox{purple}{\textcolor{white}{\small{\textsf{#1}}}}\fi}

\usepackage{listings}
\definecolor{codegreen}{rgb}{0,0.6,0}
\definecolor{codegray}{rgb}{0.5,0.5,0.5}
\definecolor{codepurple}{rgb}{0.58,0,0.82}
\definecolor{backcolour}{rgb}{0.95,0.95,0.95}

\lstdefinestyle{mystyle}{
    backgroundcolor=\color{backcolour},   
    commentstyle=\color{codepurple},
    keywordstyle=\color{NavyBlue},
    numberstyle=\tiny\color{codegray},
    stringstyle=\color{codepurple},
    basicstyle=\ttfamily\tiny,
    breakatwhitespace=true,         
    breaklines=true,                 
    captionpos=t,                    
    keepspaces=false,                 
    numbers=left,                    
    numbersep=5pt,                  
    showspaces=false,                
    showstringspaces=false,
    showtabs=false,                  
    tabsize=2
}

\AtBeginDocument{%
  \providecommand\BibTeX{{%
    \normalfont B\kern-0.5em{\scshape i\kern-0.25em b}\kern-0.8em\TeX}}}

\def\projname{LLM-for-X\xspace}
\def\llmshortcut{\texttt{Alt\,+\,1}\xspace}
\def\cOURS{\textsc{LLM-for-X}\xspace}
\def\cGPT{\textsc{ChatGPT}\xspace}

\usepackage{fancyhdr}

\begin{document}

\title[\projname: Application-agnostic Integration of Large Language Models]%
{\projname: Application-agnostic Integration of Large \\ Language Models to Support Personal Writing Workflows}

\author{Lukas Teufelberger, Xintong Liu, Zhipeng Li, \href{https://orcid.org/0000-0003-3414-7142}{Max Moebus}, and \href{https://orcid.org/0000-0001-9655-9519}{Christian Holz}}
\affiliation{\vspace{1mm}%
  \institution{Department of Computer Science}
  \city{ETH Zürich}
  \country{Switzerland}\vspace{1mm}
}
\email{firstname.lastname@inf.ethz.ch}

\renewcommand{\shortauthors}{Teufelberger et al.}
\begin{abstract}
To enhance productivity and to streamline workflows, there is a growing trend to embed large language model (LLM) functionality into applications, from browser-based web apps to native apps that run on personal computers.
Here, we introduce \emph{LLM-for-X}, a system-wide shortcut layer that seamlessly augments any application with LLM services through a lightweight popup dialog.
Our native layer seamlessly connects front-end applications to popular LLM backends, such as ChatGPT and Gemini, using their uniform chat front-ends as the programming interface or their custom API calls.
We demonstrate the benefits of LLM-for-X across a wide variety of applications, including Microsoft Office, VSCode, and Adobe Acrobat as well as popular web apps such as Overleaf.
In our evaluation, we compared LLM-for-X with ChatGPT's web interface in a series of tasks, showing that our approach can provide users with quick, efficient, and easy-to-use LLM assistance without context switching to support writing and reading tasks that is agnostic of the specific application.
\end{abstract}



\maketitle

\section{Introduction}

Large language models (LLMs) are becoming a common part of authoring processes, such as writing~\cite{design_space, saga, wordcraft, coauther}, editing~\cite{padmakumar-he-2022-machine}, and question answering~\cite{zhuang2024toolqa, NEURIPS2022_8bb0d291}.
Users across different fields now employ LLMs to enhance productivity~\cite{2023_Valentina, brynjolfsson2023generative} and creativity~\cite{the_idea_machine}.
LLM-based co-authoring and co-editing thus mark a shift in how people approach content creation~\cite{a_writers_collaborative, elephant, nichols2020collaborative} and how they perceive individual agency over the process~\cite{arnold2020predictive, gonccalves2015you, jakesch2019ai}.

The dominant mode of interaction with LLMs is chat interfaces in the form of question \& answer dialogs.
Chat assistants offered by the major providers of LLM backend services, such as OpenAI ChatGPT~\cite{chatgpt}, Google Gemini~\cite{gemini}, or Anthropic Claude~\cite{claude} mimic human conversations to make interactions with LLMs natural.
Via chat queries, users can retrieve information, conduct research, and solve their problems~\cite{pmlr-v202-gao23f, mialon2023augmented}.
These capabilities have made chat-based LLMs a promising alternative to conventional search interfaces~\cite{radlinski2017theoretical} to support users throughout task executions.

\begin{figure}[t]
  \vspace{2mm}%
  \includegraphics[width=\columnwidth]{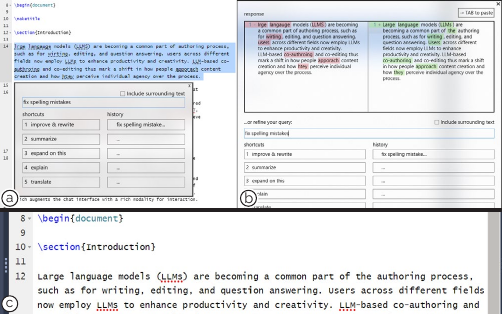}%
  \caption{
  (a)~\projname allows users to select text inside native and web apps and
  (b)~execute predefined LLM commands or enter a custom query to
  (c)~directly insert the response into the app---without the need for context switching or invoking copy \&~paste to transfer content between apps. \\[.3em]
  Video figure: \texttt{\href{https://youtube.com/watch?v=fDDMaWobjVY}{youtube.com/watch?v=fDDMaWobjVY}}}\vspace{-1mm}
  \label{fig:teaser}
\end{figure}

Beyond question \& answer interaction, users frequently copy \& paste content into and from these chat interfaces to effectively execute a variety of tasks inside \emph{other} applications.
This workflow uses the clipboard as the key interface to transfer textual information back and forth across tabs, windows, and applications.
The latest generation of LLMs also supports receiving images as input~\cite{yin2023survey_multimodal_llm} and generating images as output~\cite{dalle3}, which augments the chat interface with a rich modality for interaction.

Following the success, many developers have incorporated LLM services into their products, from smaller apps (e.g., Grammarly~\cite{grammarly}, personal AI Pi by Inflection~\cite{inflection}, My AI by Snapchat~\cite{snapchat_myai}) to large software suites (e.g., Copilot in Microsoft Office~\cite{windows_copilot}, Google Workspace~\cite{google_workspace}, Amazon Lex~\cite{amazon_lex}), often charging (separately) for this added functionality.
As a consequence, users must join multiple subscriptions, one for each app to enjoy LLM-based services---despite the large overlap of their capabilities and in the face of free general purpose use~\cite{chatgpt_free}.
For this reason, copy \& paste may remain the dominant interface for users to bridge LLM assistance and the apps they use to complete tasks in a desktop environment.

We introduce \emph{\projname}, a user interface technique that provides interaction with LLM backends to users within any frontend app.
As shown in Figure~\ref{fig:teaser}, \projname requires no copy \& paste or switching windows to LLM frontends, and it allows users to almost directly operate on text inside an app using a light-weight on-demand popup UI.
Our system service connects text selections and user queries to an LLM by either emulating user input to a chat interface (e.g., an existing subscription for ChatGPT) and seamlessly transfers responses back into the app.
Alternatively, \projname connects to LLM backend APIs to retrieve responses.
In both cases, our approach minimizes the effort to interface wiht LLMs, allowing users to focus on a task within an app without context switching.

\subsection{Walk-through and interaction design}

\projname is designed for efficient keyboard use, supporting the writing and editing process of text across apps with shortcuts.
Figure~\ref{fig:teaser} shows a typical app scenario.
The user selects text inside an app, here Overleaf running in the web browser, and triggers \projname via keyboard shortcut (\llmshortcut).
A prompt menu appears with several options for quick actions and associated prompts that can be triggered using the number keys.
Alternatively, the user may enter a custom LLM query.
Here, the user types ``fix spelling mistakes'', which sends the prompt and the text selection to an LLM chat in the background.
\projname then previews the LLM response in the menu, which the user inserts into Overleaf by pressing \texttt{TAB}.
This closes the menu and completes the interaction flow.

Below, we describe more interaction scenarios.
We also refer the reader to our video figure for additional demonstrations: \\
\texttt{\href{https://youtube.com/watch?v=fDDMaWobjVY}{https://youtube.com/watch?v=fDDMaWobjVY}}.

\begin{figure}
    \centering
    \includegraphics{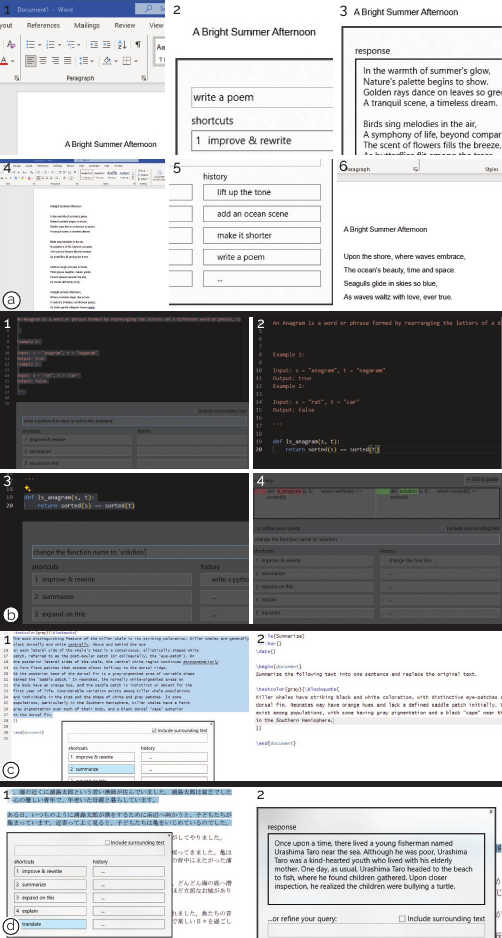}
    \caption{\projname walk-through.
    (a)~Iterating on LLM responses, 
    (b)~pasting responses as `insert below' vs. 'replacing' with diff view,
    (c)~direct in-place pasting without preview, and
    (d)~selecting and querying for information retrieval.
    \vspace{-5mm}}
    \label{fig:walkthrough}
\end{figure}

\subsubsection*{Iterating on LLM responses}

Before inserting the LLM response, \projname allows users to refine the prompt to iterate on the generated responses (Figure~\ref{fig:walkthrough}a).
Here, the dialog shows the original text and the LLM response side-by-side, highlighting changes using a diff view.
Users may alter the prompt, resubmit the query on the selected text, and see the updated LLM response in the preview area.
Our tool thereby maintains the chat session in the background, such that query refinements preserve previous output in the LLM chat interface for context.

\subsubsection*{Insert vs. replace text in the app}

While text replacement may be the most common action, \projname additionally supports appending the LLM response below the text selection when the user holds a modifier key \texttt{SHIFT} before pressing \texttt{TAB} for insertion (Figure~\ref{fig:walkthrough}b).
This functionality can be useful when the purpose of the selected text is to act as specific context to inform the completion of the task.
Examples include coding tasks where instructions or descriptions such as those in comments above a function signature should be left intact.

\subsubsection*{Additional context to the LLM}

Our tool is able to, if selected by the user, provide additional context to the LLM query by including the textual content surrounding the selection.
The additional context is flagged as context in \projname's prompt to the LLM's chat interface to ensure that the response does not directly include part of this augmentation.

In any case, \projname extracts the current app's name and the window title as context for prompt customization.
We also use the window title to decide between resuming a previous LLM chat session or starting a new session.

\subsubsection*{Direct in-place response}

When the user holds the modifier \texttt{SHIFT} while submitting a query, the prompt menu will close immediately and our system service will insert the LLM response directly in the foreground app by simulating key presses---letter by letter and in real-time as produced by the chat interface (Figure~\ref{fig:walkthrough}c).
Because this insertion appears to the foreground app as entered by the user, the LLM response can be handled as such and, for example, be undone using the app's regular `undo' function.

\subsubsection*{Querying information without inserting responses}

Finally, \projname can also be used to simply obtain additional information.
Selecting the relevant text, pressing \llmshortcut, and hitting \texttt{4} (for `explain') or \texttt{5} (for `translate') will produce the LLM's chat response in the menu's preview area as before.
This functionality can be useful for contextual information, such as in a reading app or PDF viewer when users may require context (Figure~\ref{fig:walkthrough}d).
Another example is the use of the tool for simultaneous translation (Figure~\ref{fig:walkthrough}d) and explanation on foreign language (news) websites. 

Since the app in focus may not support text entry, such as in the case of the PDF viewer, \projname recognizes this and hides the \texttt{TAB} button.
Users may still refine the query for further details.

\subsection{Design objectives}

 We primarily designed \projname to allow for quick interaction with LLM backends without the need for context (or app) switching as shown in Figure~\ref{fig:architecture}.
While our interaction design is optimized for keyboard use (e.g., shortcuts to summon via \llmshortcut, to select menu items via 1..5, \texttt{TAB} for pasting, and modifier keys), all options are equally accessible via mouse use, too.
Inside native apps, \projname's prompt menu can be invoked using an icon in the system tray, whereas a toolbar icon triggers it in the browser.
The menu stays open and in front until the user presses \texttt{ESC} or clicks the button in the top right. 
Upon pressing \texttt{TAB}, \projname inserts or replaces the selection at the location of the cursor.
Our menu will remain active even if users click a different position in the text editor, such that they can determine where to paste the LLM response before pressing \texttt{TAB} or \texttt{ESC} to close the menu.

The visual design if \projname's popup is lightweight to minimize distraction from the foreground app.
\projname's UI has three main elements: query input field (which is in focus when the menu appears), suggested actions (chosen based on popular use-cases of ChatGPT~\cite{understanding_user}, accessible via shortcuts), and the preview output field for LLM responses, where changes are color-highlighted.
The latter can be navigated with \texttt{Page Up}/\texttt{Page Down} or the scrollbar, while typing will modify the query prompt.

\subsection{Contributions}

Our research makes two main contributions:

\begin{itemize}[leftmargin=*]

\item
\projname, an operating system-wide shortcut interaction technique that allows prompting LLM backends from within any native app or web app, optionally based on a text selection for manipulation.
Our technique is implemented through an OS service and can alternatively interface with a browser extension to facilitate LLM communication or an LLM API backend directly, either offered by popular LLM providers such as OpenAI ChatGPT and Google Gemini.
Our technique thus bridges their text-based functionality with text input inside \emph{any} app.

\item a user study with 14 participants to compare their performance during writing, editing, and coding tasks using \cOURS and \cGPT.
Participants were significantly faster in editing text and reported higher usability (using SUS~\cite{SUS}), as \projname allowed them to stay within one app without context switching.

\end{itemize}

\section{Related work}

\projname is related to virtual assistants, LLM-based support tools, and shortcut interaction.

\subsection{Virtual Assistants}

A virtual assistant (VA) is a software agent that carries out a variety of tasks or services in response to user input, such as commands, questions, and verbal prompts~\cite{va}. 
\citeauthor{design_space} categorize the research of intelligent writing assistants into task, user, technology, interaction, and ecosystems~\cite{design_space}.
\projname falls under the interaction aspect, specifically emphasizing the hybrid interaction metaphor, localized user interface, and explicit user control. 

\subsubsection*{Intelligent Personal Assistants}
With the integration of machine learning and advanced speech recognition technologies into personal devices, the prevalence of intelligent personal assistants (IPAs) highlights the effectiveness of AI-powered personal assistants in everyday tasks~\cite{ipa_overview}. 
These software agents are present in wearables, smartphones, and desktops~\cite{ipa_overview}, such as Siri~\cite{siri}, Alexa~\cite{Alexa}, and Cortana~\cite{cortana}.
However, most users rely on IPAs for only basic tasks, such as playing music and scheduling appointments~\cite{dubiel2018survey, cowan2017can}.

\subsubsection*{IPAs in the Age of LLMs}
Researchers envision the next generation of IPAs to be powered by LLMs and to become more prevalent as omnipresent personal companions~\cite{dong2023towards}. 
Early prototypes for LLM-based IPAs on smartphones include GPTVoiceTasker~\cite{vu2024gptvoicetasker} and GPTDroid~\cite{liu2023chatting}. 
Windows Copilot~\cite{windows_copilot} is among the first commercially-available personal companions for desktop environments. 

\begin{figure}[t]
  \vspace{1mm}%
  \includegraphics[width=\columnwidth]{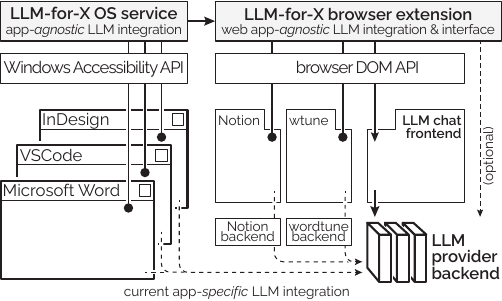}%
  \caption{\projname offers a system-wide shortcut from any web or native application to various text-based LLM backend.}
  \label{fig:architecture}
\end{figure}

\subsection{LLM-based Tools}
With the rapid expansion of open-source NLP research, there are many studies on specialized bots and apps for specific use cases such as domain-specific or task-orientated text generation~\cite{gpts,decoding_chatgpt}. 
Popular uses for LLMs include text generation, information retrieval, and problem solving~\cite{understanding_user,decoding_chatgpt}. 

\subsubsection*{Writing}
Co-writing systems such as Wordcraft~\cite{wordcraft} show that users make use of AI assistants throughout many stages in the writing process~\cite{elephant} and tools such as or TexGPT from Writeful~\cite{writeful} even offer platforms-specific AI assistants, e.g., for Overleaf.
More advanced writing assistants focus on providing new perspectives for reflection and inspiration in addition to grammar correction and text polishing~\cite{beyond_text_generation,the_idea_machine}. 

\subsubsection*{Information Retrieval}
Learning and browsing online could also benefit from assistance.
Complex search tasks require similarly complex search prompts~\cite{hassan2014supporting}.
The success of complex search tasks is then heavily dependent on the quality of the search prompt~\cite{white2023navigating}.
GPT models also possess suitable translation capabilities~\cite{translation} and can be useful in educational settings~\cite{chatgpt_for_good, educsci13070692}.

\subsubsection*{Coding}
AI assistants are also popular for aiding software engineering tasks as so-called copilots or companions~\cite{ross2023programmer}.
The advancements of LLMs fine-tuned on publicly available code facilitate more efficient coding with AI assistants (e.g., Codex~\cite{chen2021evaluating_llm_code}).
Previous studies showed that while Github Copilot and ChatGPT-assisted coding does not outperform human-written solutions, users acknowledge their ability to handle fundamental tasks~\cite{expectation_vs_experience, adamson2023assessing}.

\subsection{Efficient interaction and shortcuts}
When trying to harness the full power of LLMs and AI assistants, the bottleneck during interaction with generative AI models is speed and output quality~\cite{strobelt2022interactive,mishra2023promptaid}.
Since many LLM commands are invoked through menus, keyboards can provide preferable and faster access over mouse use~\cite{mouse_keyboard} via shortcuts.

Further, using AI assistants to everyday tasks should disrupt the underlying tasks as little as possible~\cite{zide2017work}.
Across a broad range of professions including primary care~\cite{chisholm2001work} and software engineering~\cite{leroy2020interruptions}, such interruptions otherwise significantly impair productivity and quality of work, even those as small as context switching.
For software engineers, context switching strongly affects perceived productivity and mood~\cite{meyer2017work}.
AI assistants should thus be seamlessly integrated into everyday tasks to minimize interruptions and context switches.

\section{\projname Implementation}

Figure~\ref{fig:architecture} shows an overview of the architecture of our implementation, including \projname's operating system (OS)-level background service, which produces its prompt menu UI, its browser extension, and its optional direct API use.
\projname's prompt menu is the main interactive and user-facing element, configured by our background service that monitors for short-cut activation.
This service also interfaces with the content inside native apps and retrieves surrounding content for context.
For interfacing with web apps, our system includes a browser extension for deep access to web pages.
In both cases, our background service retrieves selected text and surrounding content and replaces or inserts selections with the responses from the LLM.
Our background service also interfaces with the web-based frontend of LLM services via our browser extension and operates them by simulating a user's copy\&paste interaction or alternatively directly queries backends via their APIs.

We implemented \projname to run on Windows\,10 and up.
It supports Google Chrome or Microsoft Edge through its extension API.
If support for native apps is not needed, \projname can also operate solely based on its browser extension (including an in-browser short-cut listener and a browser-based menu).
This makes it compatible with all platforms that fully support Chrome or Edge and extensions without requiring a native component.

\subsection{OS-level background service}

\projname runs as a background app to listen for triggers of the prompt menu across applications and web browser tabs.
We implemented the service is C\# using .net APIs, registering a system-level keyboard hook to watch for global keyboard shortcuts.
Once our service detects a shortcut, our native app interface extracts application details and text selection to prepare the prompt.

\subsection{Native app interface (via accessibility API)}

Access to text contents within an app is available via UI Automation API (UIA) on Windows, which is part of the Microsoft Accessibility API.
Our service has a registry key in Windows set to be able to communicate with the browser. 
From the currently focused element, we obtain the window title and Process ID (PID), the latter we use to capture the current app's name.
Both window title and app name we use for context. 

To extract selected text, we start with the focused element and browse the subtree of elements for a UI element with a UIA's \texttt{TextPattern}.
Once the property is available, we discover the content for the prompt body via UIA's \texttt{GetSelection}.

\projname can also obtain surrounding text for additional context as mentioned above.
We reuse the identifier of the UI element with the \texttt{TextPattern} and obtain the entire body of the text.
While we use the selected text as main input into the prompt, the rest we add to the prompt for context explicitly declared as such.

\subsubsection*{Paste LLM responses}
Finally, to insert the responses from the chat interface, we emulate direct user input into the control.
\projname's backend first copies the response to the clipboard and then invokes `Paste' with \texttt{SendKeys.SendWait} to simulate the paste operation.
The benefit of proceeding this way is that input gets entered into the app's undo/redo stack.
In case the text is not to be replaced but inserted below, we simulate a `Cursor Right' key press beforehand.

\subsubsection{Fallback: if the API fails}

While the UI Automation API supports most apps on Windows, we have found some not to reveal text content or selections.
In case \projname discovers no existence of \texttt{TextPattern}, our service makes use of the clipboard and injects a copy shortcut command (\texttt{CTRL\,+\,C}).

Likewise, to retrieve the surrounding context if so configured, we simulate `Select All' (\texttt{CTRL\,+\,A}) and another copy shortcut.
Just before, we simulate pressing `Cursor Right' and `Space' to enter an operation to the undo stack.
After selecting all and copying the text, we undo the last command (space) to set the cursor position to the bottom of the selection.
We cannot recover the text selection state, however.

\subsection{Web app interface (via browser extension)}

If a web browser is in the foreground, \projname's browser extension connects to our backend service upon detecting the trigger shortcut.
The extension then interfaces with the foreground web page to extract the selected text (if any), textual content, and later handle text replacement or insertion.

Through Javascript and the DOM API, \projname's browser extension obtains the selected element in the DOM and extracts the selected text inside.
Our extension then finds the first common ancestor and takes its \texttt{parentNode} to extract the selection's context.

\subsubsection*{Insert LLM responses}
\projname's extension again uses the current selection and appends \texttt{textNodes} to the end. 
This reliably works for both editable UI elements (e.g., \texttt{textarea} and \texttt{input}) but also any other elements (e.g., \texttt{div} with \texttt{contenteditable} set).

\subsubsection*{If responses cannot be pasted}

Finally, our extension determines whether the element that contains the selected text is not editable, such that an LLM response could not be pasted.
In this case, \projname's prompt menu does not show the \texttt{TAB} button.
We conclude this from the element's tag name (input or textarea) and from whether it is set to read-only, disabled, or if it is \texttt{contenteditable}.

\subsection{Native service-to-extension interface}

For communication between our browser extension and native service, we send commands via the Native Messaging API~\cite{ChromeDocsNativeMessaging}.
Our browser extension thereby starts the native app and uses the standard input/output interface for message exchange.

\subsection{LLM interface (via browser extension)}
\projname currently supports ChatGPT, Mistral, and Gemini.
Because they all follow the same interaction paradigm using a chat interface, our browser extension emulates user input when a query is submitted from the prompt menu, extracts the response from the LLM web UI, and transfers it back to the prompt menu.

\subsubsection{Communicating with LLM chat interface pages}

Upon opening an LLM provider's chat interface, our extension detects the chat window, the query input field, and the response element.
The extension searches for the input element using a combination of tag names and element paths.
It then inserts the query by either setting the \texttt{textContent} or \texttt{value} of the element with dispatching a few events first to simulate user interaction.
The query is submitted by injecting a `return' or `click' event.

Afterwards, our extension repeatedly polls the DOM in short intervals to detect changes in the output element.
We obtain the text from the \texttt{innerText} property of the output element.
Whenever a change in content is detected, our extension forwards the complete content to our OS-level service and into the prompt menu's preview field.
Once no changes have been detected for a predefined duration, it will signal to our background service that the response is complete.
We opted for this procedure, because chat interfaces perform various operations on the returned raw response throughout, such as replacing symbols.

\subsubsection*{LLM chat interfaces and changes}

\projname has default tag names and element paths configured for the currently supported chat interfaces of ChatGPT, Gemini, and Mistral.
In the case that a chat interface changes, our extension  first uses heuristics to rediscover the elements (looking for the \texttt{tagnames}, \texttt{id}, and an overlap in \texttt{classes} added, or the same path in the DOM tree.
If this fails, our extension reports a problem and provide a tool to select the input box and the output element manually (similar to Chrome's `Select an element to inspect' function) to update our settings.
Because all parameters and URLs are stored in the extension settings, users could also add further LLM chat interfaces in the future.

\subsubsection*{Browser handling to optimize for responsiveness}

When a user triggers \projname's prompt menu, our browser extension opens the LLM's web page in a browser background window.
If the user has interacted with the current foreground app with this title before using \projname, our browser extension loads the corresponding previous chat, such that the user can continue prompting within the context of the previous interaction sequence.
This accounts for loading times while the user still sees and operates \projname's prompt menu, ensuring that upon submitting a query our tool does not add to the response latency.

\subsubsection{Prompt padding}

For submitting the prompt entered in \projname's prompt menu (or that associated with a menu item), we augment the text selection as follows:
\begin{lstlisting}
Your task is to answer the following query from the user. Do not express approval or your own judgment of the query. Just respond with a clear answer. If prompted for code, just output the code, no explanation, just one response of code and nothing else.

User query: %USER_QUERY%

The user's query refers to this specific text:
%SELECTED_TEXT%

The user issued the query while working with %APP_NAME% and the document %WINDOW_TITLE%.

The user has provided additional context for their query. Do not directly quote this context, but use it to formulate a response.

Context: %TEXT_CONTEXT%
\end{lstlisting}

\subsection{Direct LLM interface via API (optional)}

As an alternative to browser-based LLM access, \projname also supports direct interaction with LLM backends via their APIs.
While this incurs query-specific charges for each use of \projname, it affords more freedom over query types while still bringing LLM support to many apps.
Responses are also faster than those from the browser-based interface without variable-speed delays.

\section{User study}

The purpose of our study was to evaluate the efficacy of \projname in the context of authoring, reading, and coding tasks.
In a controlled user study, we compared participants' performance solving these tasks using one of two \emph{Interfaces}, \cOURS and \cGPT (version 3.5 through the web interface).

\subsection{Tasks}

Participants completed three tasks during the study.
Each task had two instantiations that were similar in difficulty and format but varied in specific content.

\subsubsection{Writing}

The writing task comprised three common subtasks for intelligent writing assistants: summarizing, editing, and text composing~\cite{the_idea_machine, beyond_text_generation, dimensions_for_designing}.
For summarizing, participants summarized a paragraph from an academic paper draft in Overleaf.
For editing, they rewrote a narrative paragraph in Microsoft Word in a different style.
For composing, they wrote an email in Microsoft Outlook given a use-case: an invitation to an after-work celebration and an out-of-office announcement for a personal vacation.

\subsubsection{Reading}

For the purpose of answering a question, participants read a one page long folk story in a foreign language that they were not familiar with in Acrobat Reader.
Their task was to answer a question about the text in English.

\subsubsection{Coding}

Participants completed a simple coding task using Python in VSCode similar to previous studies~\cite{expectation_vs_experience, adamson2023assessing}.
The task involved extracting information from a CSV file, aggregating dataframes, and writing to a text file as output.

\subsection{Procedure}

The study started with a general introduction to the study context and explained the tasks to complete.
Participants received an introduction to and demonstration of \cGPT and \cOURS, and they were allowed to train in both for up to 20\,minutes, using any native or web apps they wanted.

Participants then moved on to the main part of the study, for which they received written instructions.
Participants completed all three tasks using the same condition, took a break, and repeated the tasks in the second instantiation with the other condition.
The conditions \cOURS and \cGPT were counterbalanced across participants.
During each (sub)task, a timer on the screen displayed the remaining time to participants.

After completing all tasks using one interface, participants filled out a questionnaire that included SUS~\cite{SUS} and NASA TLX~\cite{tlx}.
When participants had finished both conditions, they had the option to provide qualitative feedback about both conditions.

The study ran on a Windows\,10 computer with a full keyboard, mouse, and a 27\textquotedbl\ display.
An experimenter monitored participants' performance, answered questions, and ensured correct task outcomes.
All participants completed the study in less than an hour.

\subsection{Participants}

We recruited 14 participants from various departments within our institution (5 female, 9 male, ages 22--37 years).
All participants had prior experience with Python. 
More than half of them reported that they use ChatGPT and other LLM-based tools (e.g., Copilot) multiple times a day. 
Coding is the most popular reported task for LLM use, and 12 participants that they routinely use ChatGPT for coding.
9 mentioned using it for writing, editing, and conducting research.
7 reported using ChatGPT for drafting emails and messages. 
Participants received a small gratuity for their time after the study.

\subsection{Measurements}

We recorded the following metrics to assess participants' performance and impressions during the study:
(sub)task completion time, time-stamped logs during task completion, Likert-scale questionnaire responses, and qualitative comments.

\subsubsection{Completion time}

The timer started once participants had completed reading the instructions of each task and had asked questions if they had any.
The timer stopped only when the experimenter had verified the correct completion of the task.
For writing, the text needed to be complete with no placeholders remaining (as \cGPT often includes them).
For reading and question answering, the answer needed to be correct, otherwise participants could try again.
For coding, their code needed to produce the correct output, for which we provided a verification script that they ran on the command line.
We measure all completion times in seconds.

\subsubsection{System Usability Scale}

We assessed the usability of \cOURS and \cGPT for daily desktop-based tasks through the system usability scale (SUS)~\cite{SUS}.
In particular, we scored the outcome of the SUS using the following standard formula~\cite{lewis2018system}:
$2.5 \ast ( 20 + \sum_{i \in 1,3,5,7,9}{SUS_i}  - \sum_{i \in 2,4,6,8,10}{SUS_i} ),$
where $SUS_i$ denotes the $i^{th}$ question in the SUS for $i\in 1, \ldots,10$.

\subsubsection{Task Load Index}

We used the NASA task load index (NASA TLX) questionnaire~\cite{tlx} to assess participants' perceived workload while using LLM interfaces.
It consists on six sub-scales on mental demand, physical demand, temporal demand, performance, effort, and frustration level.
To increase the convenience for study participants, we adapted each sub-scale to a 5-point Likert item ranging from `very low' to `very high'.
We then calculated the load index via the following formula, which mimics the standard aggregation of the NASA TLX~\cite{tlx}: $TLX = \frac{25}{6} \sum_{i\in 1, \ldots, 6}{(TLX_i -1)},$ where $TLX_i$ is the $i^{th}$ sub-scale of our adapted NASA TLX questionnaire.

\subsubsection{Questionnaire}

In addition to SUS and NASA TLX, participants provided their experience in using LLM-based support for the tasks (following \citeauthor{adamson2023assessing}'s study~\cite{adamson2023assessing}). 
A 5-point Likert scale assessed the condition's perceived impact on their efficiency and productivity, how often they needed to edit the output content to meet task requirements, and the follow-up work needed to reformat the output to produce the final result.

\subsubsection{Feedback}

After the study, participants could report any feedback they had on either interface.
Participants commented on the task design, UI design, and the experience of solving the tasks using \cOURS and \cGPT. 
After their initial comments, we followed up with questions as reported in previous work~\cite{adamson2023assessing}.

\subsection{Results}

\subsubsection{Completion time}

\begin{figure}[t]
    \centering
    \includegraphics[width=\columnwidth]{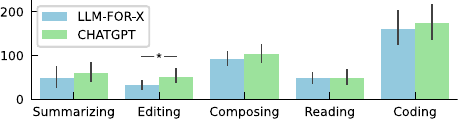}
    \caption{Effect of \emph{Interface} on task completion time [sec].}
    \label{fig:completion_time}
\end{figure}

Figure~\ref{fig:completion_time} shows all completion times.
We could not find a statistically significant difference of total completion time across all tasks (\cOURS: $M=382.21$, $SD=127.09$; \cGPT: $M=436.07$, $SD=167.19$; \wilcoxon{80.5}{=}{0.43}).
We did find a main effect of \emph{Interface} on \emph{Completion Time} during the editing task (\wilcoxon{-2.18}{<}{0.05}), where participants completed the task significantly faster using \cOURS ($M=31.71$, $SD=19.04$) than when using \cGPT ($M=51.14$, $SD=32.50$). 
We did not observe any other significant differences.

\subsubsection{System Usability Scale}

\begin{figure}[t]
    \centering
    \includegraphics[width=\columnwidth]{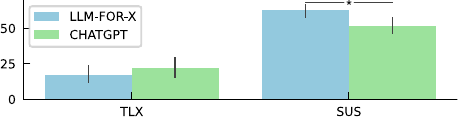}
    \caption{Effect of \emph{Interface} on SUS and NASA TLX scores.}
    \label{fig:sus}
\end{figure}

We found a significant effect of \emph{Interface} on \emph{System Usability} across all tasks (\wilcoxon{2.64}{<}{0.01}).
Post-hoc tests showed that participants rated \cOURS ($M=62.54$, $SD=9.2$) significantly higher than \cGPT ($M=51.68$, $SD=10.94$) (\autoref{fig:sus}).

\subsubsection{Task Load Index}

We did not find a statistically significant overall difference for NASA TLX (\wilcoxon{-1.03}{=}{0.30}).
Participants on average scored \cGPT with 22.32 ($SD=14.58$) and \cOURS with 17.26 ($SD=11.77$).
However, participants rated \cOURS significantly lower on the effort-related sub-scale of the NASA TLX for ease of use ($M=1.64$, $SD=0.49$) compared to \cGPT ($M=2.27$, $SD=1.03$).
We found no other significant difference between sub-scales of the NASA TLX.

\subsubsection{Questionnaire}

We could not find a statistically significant difference in the questionnaire answers.
Participants' reported answers for efficiency were \cOURS: $M=4.64$ ($SD=0.63$) vs. \cGPT: $M=4.50$ ($SD=0.65$).
For perceived quality of their work, they rated \cOURS with $M=4.43$ ($SD=0.65$) and \cGPT with $M=4.21$ ($SD=0.89$).
On perceived difficulty, they rated \cOURS with $M=1.71$ ($SD=0.61$) compared to \cGPT ($M=1.57, SD=0.65$).
For number of needed modifications they needed to perform, they rated \cOURS with $M=2.43$ ($SD=0.51$) and \cGPT with  $M=3.07$ ($SD=1.21$) and for ``less follow-up work was required'' they scored \cOURS: $M=2.21$ ($SD=0.80$) vs. \cGPT: $M=2.64$ ($SD=1.28$).

\subsubsection{Feedback}

Participants expressed that they felt more efficient when conducting the tasks with \cOURS, because it eliminates the need for context switching. 
P1 said \textit{``Because \cOURS is more integrated into the environment, I felt less the need to switch my focus compared to when I use \cGPT.''} 
The shortcuts for menu initiation and insertion were also appreciated. 
For example, P7 said \textit{``It is cool that no copy paste is required when using the tool.''}

However, some participants still prefer the \cGPT interface for user-friendliness. 
P6 pointed out the personalization of \cGPT saying \textit{``\cGPT talks in a more friendly manner and feels more like a personal assistant.''}. 
P2 found the code generated by the \cGPT web app easier to read.

\section{Discussion}

Overall, the results of our evaluation were encouraging for our approach.
While \cOURS achieved a significant difference in performance only for the editing task, interestingly with a 40\% faster completion time on average, participants' ratings on the SUS and for ease-of-use on TLX confirmed our original motivation.
It is worth pointing out that all participants were at least well familiar with \cGPT and that the majority of them used it frequently, such that we can assume that they were trained to the point of routine operation.
Therefore, \cGPT was a challenging baseline despite the training participants received for \cOURS.

One observation during the study was that participants' use of \cGPT varied based on personal preference.
Some heavily relied on it, copying \& pasting in large amounts of contents, while others use them more deliberately and sparingly.
For example, when replacing the placeholders in the composed emails, more than half of the participants insisted and requested in a prompt that this be done by the LLM, while the minority just filled them out manually.

During or after the study, no participant commented on the choice of shortcuts or other design choices of \cOURS, except for the 2 who commented on the lack of a `personalized assistant experience'.
Taken together, we interpret this as an encouraging sign that our design objectives for the UI and the interaction did not interfere with participants' workflows.
On the contrary, participants offered several recommendations how the UI could be extended to better support the tasks at hand.
A frequent suggestion was that the menu buttons (1..5) should be generated based on the context of the selected text and application window.

\section{Limitations and future work}

\subsubsection*{In-situ editing}

Participants took time to get used to \projname's UI, as LLM responses flowed directly into the foreground app upon pressing \texttt{TAB}.
Despite the preview in our UI, they were used to seeing the formatted output in the context of previous responses and using copy \& paste to transfer (portions of) the response.
Future implementations could include collapsible displays of history for previously generated responses, syntax highlighting, or the option to insert only portions of the previewed response.

\subsubsection*{\projname vs. apps with built-in LLM support}

While our approach provides general-purpose text generation and editing across apps, those with built-in LLM assistance can offer better tailored support.
For example, Github Copilot's Codex is specifically integrated into programming environments (e.g., VSCode, JetBrains).
Similarly, assistants in creativity apps can generate and manipulate app-specific objects (e.g., Adobe Illustrator or Premiere), where mere text responses are insufficient.

\subsubsection*{Context is limited to text}

Our current implementation is limited to sourcing text from the surrounding context.
Future work could additionally include screenshots to GPT-4, such as to capture a screenshot of the rendered draft when editing in Overleaf or capturing surrounding images in InDesign for suggestions.

\subsubsection*{Integration with other web-based query services}

\projname's focus is on LLM-based chat interactions, but our implementation can also interface with other web-based services, such as Google Translate and DeepL (see video), as query elements and response fields are equally well defined on the web page.
Future implementations could explore more general media types (e.g., Wolfram Alpha).

\section{Conclusion}

We have presented \projname, a system-wide bridge between applications and LLM-based assistance for textual content.
By pressing a shortcut, our system shows a lightweight overlay to query LLM chat systems with a text selection and allows the response to seamlessly transfer back into the app.
In the background, our system interfaces with browser-based LLM chat interfaces to submit the query and process the response or it can query LLM backends directly through their APIs.
\projname thus allows users to leverage LLM capabilities for text manipulation across virtually all apps, even if an app does not build in native LLM support.

In our comparison of \projname with ChatGPT, we found statistically significant differences only in its ability to support text editing more quickly as part of a rewrite task and its perceived usability (rated on SUS~\cite{SUS}).
Given these results and the qualitative insights during the study as the basis for future developments, we believe that \projname is a promising option for supporting text manipulation tasks and supporting quality of work across applications, without investing in app-specific LLM subscriptions.

\balance
\bibliographystyle{_templates/ACM-Reference-Format}
\bibliography{llm4x}

\end{document}